# Cosmo-physical effects in structure of the daily and yearly periods of change in the shape of the histograms constructed by results of measurements of alpha-activity $^{239}$Pu.

S.E.Shnol[1,2]  K.I.Zenchenko[2] , N.V.Udaltsova[2]

[1] **Physical faculty of the Moscow State University**

[2] **Institute of Theoretical and Experimental Biophysics of the Russian Academy of Science**

## Introduction (Summary)

As shown in the previous publications, the shape of the histograms constructed by measurements of alpha-activity of samples of $^{239}$Pu, changes with periods equal approximately 24 hours, 27 days, and a year. At a higher resolution each of these periods splits as a minimum in two components: daily period consists of two: sidereal day (1436 minutes) and solar day (1440 minutes), 27-days period splits in 2 to 3 "sub-periods", and the yearly period appears to be a join of "calendar" period (which is equal to 1440 x 365 = 525600 minutes) and "solar" period (equal to [(1440 x 365) + 369] =525969 minutes) [4].

In the present paper results of more detailed research of this phenomenon are offered.

## Materials and methods

Initial materials in our research are results of measurements of alpha-activity of samples $^{239}$Pu by semi-conductor detectors. Measurements are conducted with one-second intervals round the clock, during many years [1-4]. We build histograms by results of measurements and estimate similarity of histogram shapes. The detailed description of methods of measurements, construction of histograms and comparison of histogram shapes, and also criteria of an estimation of statistical reliability of obtained results are published in [1-4,8,9]. Histograms built either by 60 results of one-second measurements –"one-minute histograms", or by 60 results of one-minute measurements – "hourly histograms".

The basic object in the present research is dependence of likelihood of repeated occurrence of histograms of the given shape on size of time interval dividing similar histograms.

# Results

## Approximately Daily Periods

On Fig.1 the typical result, dependence of likelihood of repeated occurrence of hourly histograms of the given shape on size of time interval dividing similar histograms, is represented. On Fig.1 two major effects are visible.

1) Effect of "a near zone" – significantly higher likelihood of similarity of the form of the nearest next histograms;
2) About-daily period of repeated occurrence of histograms of the similar shape.

95 % confidential intervals for peaks at 1 and 24 hours (calculated by Poisson distribution [6]) confirm reliability of observable effects.

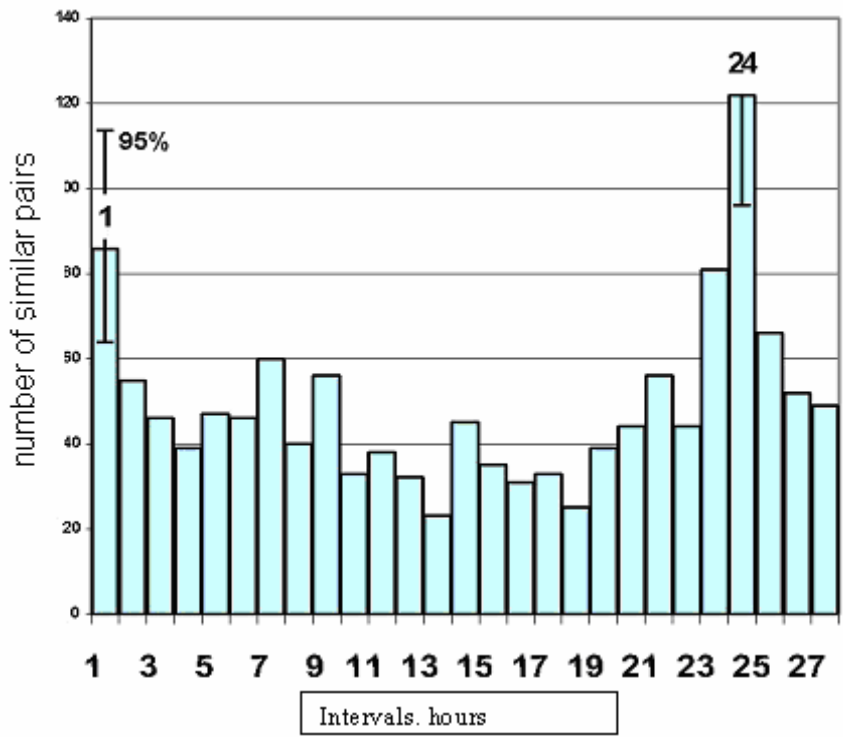

**Fig. 1. Dependence of likelihood of repeated realization of hourly histograms of the given shape on size of time interval dividing similar histograms. Measurements of alpha-activity of a sample $^{239}$Pu by detector with static collimator, directed to the West, June 8 – 30, 2004.** Abscissa is time interval in hours. Ordinate is number of similar histogram pairs. 95% confidence intervals for peaks at 1 and 24 hours are marked.

Measurements June 8 – 30, 2004 during 538 hours. 14526 pair comparisons are made. 1323 similar pairs (9,1 %) are chosen. The height of peak at 24 hour interval makes 122 pairs or about 24 % from greatest possible. A background about 8 %. Peak of "a near zone" is 86 pairs or about 16 %.

A similar distribution, though at a higher resolution – for one-minute histograms, is presented at Fig.2. Now an individual hourly interval at Fig.1 corresponds to 60 one-minute intervals at Fig. 2.

At such resolution about-daily period was divided into two components: the first is equal to 1436 minutes («sidereal day»), and the second is 1440 minutes («solar day»). Thus, the shape of histograms is defined by an exposition of the laboratory to both: the celestial sphere, and the Sun. It is necessary to note «high resolution» of our method – presence of rather narrow peaks corresponding to these two periods.

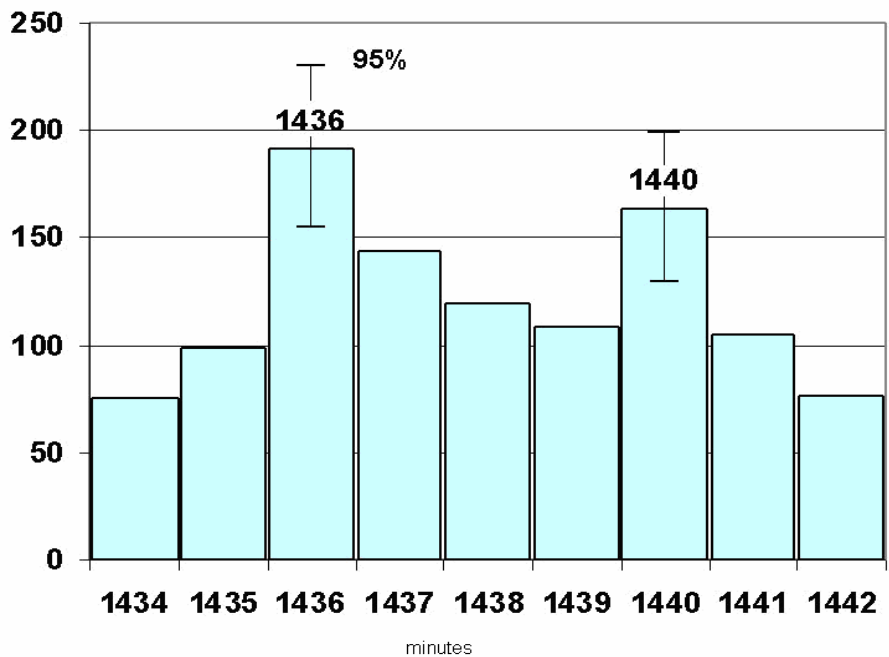

**Fig. 2. Approximately daily period of repeated realization of histograms of the given shape with the one-minute resolution. Measurements on May 29 - June 1 2004, the counter with static collimator, directed on the West.** Abscissa is time interval in minutes. Ordinate is number of similar histogram pairs. 95% confidence intervals are marked.

Results of experiments with collimators restricting a flow of the alpha particles of radioactive decay in different directions, serve as a confirmation of conclusion about dependence of histogram shapes on an exposition of laboratory to the celestial sphere and the Sun. Results of an experiment with collimator making 3 revolutions a day, counter-clockwise in a plane parallel Celestial Equator [11], are presented at Fig.3. In view of one more revolution made by the Earth daily rotation, we acquire 4 revolutions in a day.

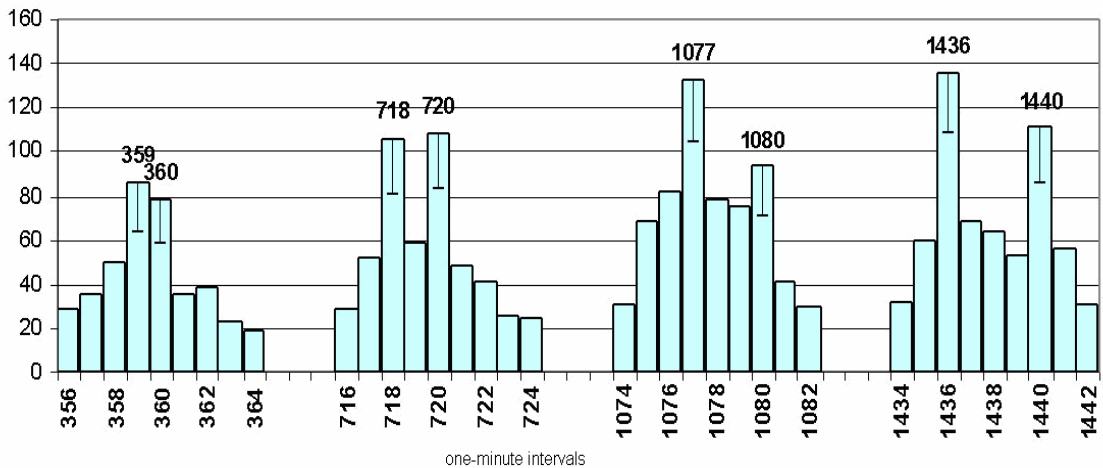

**Fig. 3. Experiments with rotated collimators. 3 revolutions counter-clockwise plus one revolution of the Earth. Occurrence of 6-hour (360 minute) periods of realization of similar one-minute histograms. With each new revolution regarding the Celestial sphere, the degree of distinction of the "sidereal" and "solar" periods grows.** Lower 95% confidence limits are marked.

**Measurements of May 29 –September 6, 2004. 1-minute histograms (by 60 one-second measurements), 7 times smoothed. Abscissa is time interval in minutes. Ordinate is number of similar histogram pairs. By 12600 comparisons of similarity of histograms in each period. Total in four = 50 400 comparisons.**

Apparently on fig. 3, with each revolution of collimator, distinction of "sidereal" and "solar" extrema grows. In the first 6-hour period two unshared extrema equal to 359 and 360 minutes are visible. At two revolutions – in the 12-hour period corresponding extrema of 718 and 720 minutes are quite resolved and divided by one interval. At three revolutions – in 18 hour period - extrema of 1077 and 1080 minutes, divided by two intervals are visible, at 4 revolutions the resolution is even stronger – extrema of 1436 and 1440 minutes, are divided by three intervals.

Thus, similar histograms are observed in process of collimators rotation, when alpha–particles flow in identical directions regarding Celestial Sphere or the Sun.

## Yearly periods.

Similar histograms with high likelihood are repeatedly realized with approximately yearly periods [1,3,4]. Detailed research of this phenomenon has led to revealing here again two precise periods – "calendar"(365 days) and "sidereal"(365 days plus 6 hours). At higher resolution when comparing one-minute histograms, the following important details come to light:

The "calendar" extremum of likelihood of occurrence of similar histograms is observed not in one year precisely, but at the period which is one minute less then number of minutes in one year (1440 x 365 = 525600), i.e. in 525599 minutes; for two years it is exactly two minutes earlier, i.e. in 1 051 198 minutes, for three years it is exactly three minutes earlier, i.e. in 1 576 797 minutes.

The "sidereal" extremum of histogram similarity is observed in 365 days and 369 minutes (6 hours plus 09 minutes). For two-year interval delay is equal 369 x 2 = to 738 minutes. In three years delay makes 369 x 3 = 1107 minutes.

Figures 4 – 9 illustrate above-mentioned phenomena.

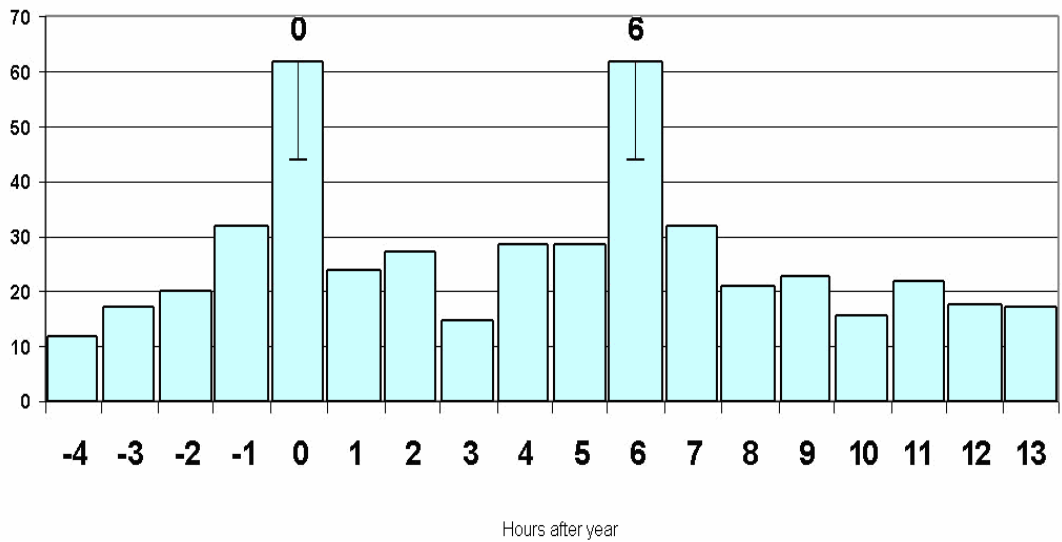

**Fig. 4. The yearly period of occurrence of similar histograms, as well as about-daily, consists of two extrema. Similar hourly histograms with high likelihood are realized equally in a year (365 days) – "the calendar period"; and in a year plus 6 hours (365,25 day) – the "sidereal" period. Distribution is obtained after comparison of the histograms constructed by results of measurements with one-year interval October 23 through November 10 in 2000 and 2001. Abscissa is time interval in hours minus one year. Ordinate is number of similar histogram pairs corresponding to the given size of an interval. Lower 95% confidence limits are marked for peaks.**

It is visible on fig. 4 that the likelihood of realization of hourly histograms of the similar form sharply grows: 1) at the same dates and hours exactly in a year – "calendar" period, and 2) in a year and 6 hours – the "sidereal" period.

When comparing one-minute histograms, it is evident (fig. 5), that the "calendar" period is shorter then one year for 1 minute: it is equal not to 525600 minutes, but 525599 minutes, and the "sidereal" period is equal to calculated value (525969 minutes)

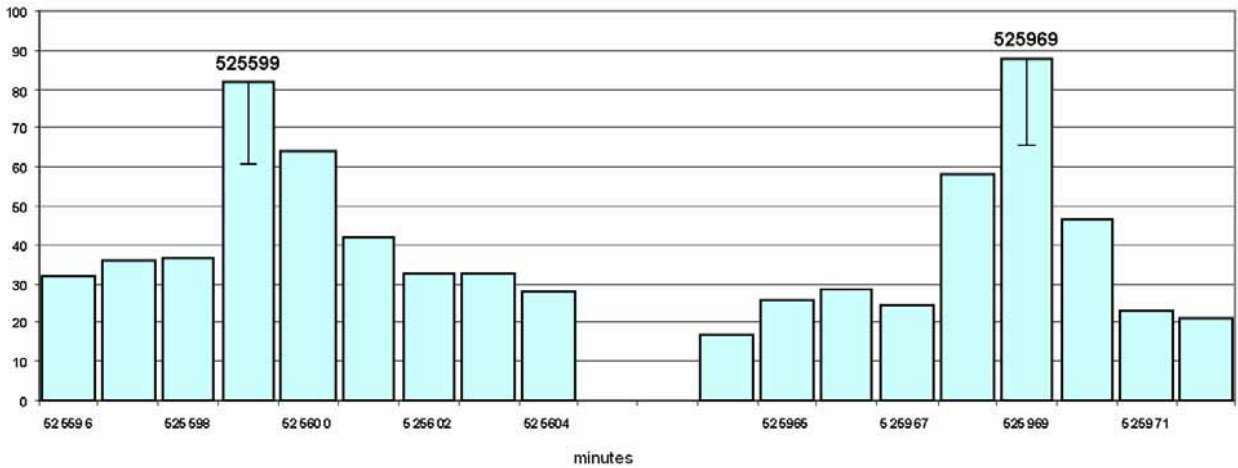

**Fig. 5. When comparing one-minute histograms "calendar" period is equal to 525599 minutes (1 minute less then number of minutes in a year), and the "sidereal" period is equal to 525969 minutes, i.e. 365 day 6 hours and 09 minutes. Measurements with one-year interval: November, 24, 2001 and November, 24, 2002. Abscissa is time interval in minutes. Ordinate is number of similar histogram pairs corresponding to the given size of an interval. Lower 95% confidence limits are marked for peaks.**

One-minute difference of "calendar" period from exactly one year has been confirmed in large series of experiments with comparison of 80 000 pairs of histograms.

Fig. 6 represents results of comparison of hourly histograms constructed by measurements with two-year interval in August – September 2000 and 2002.

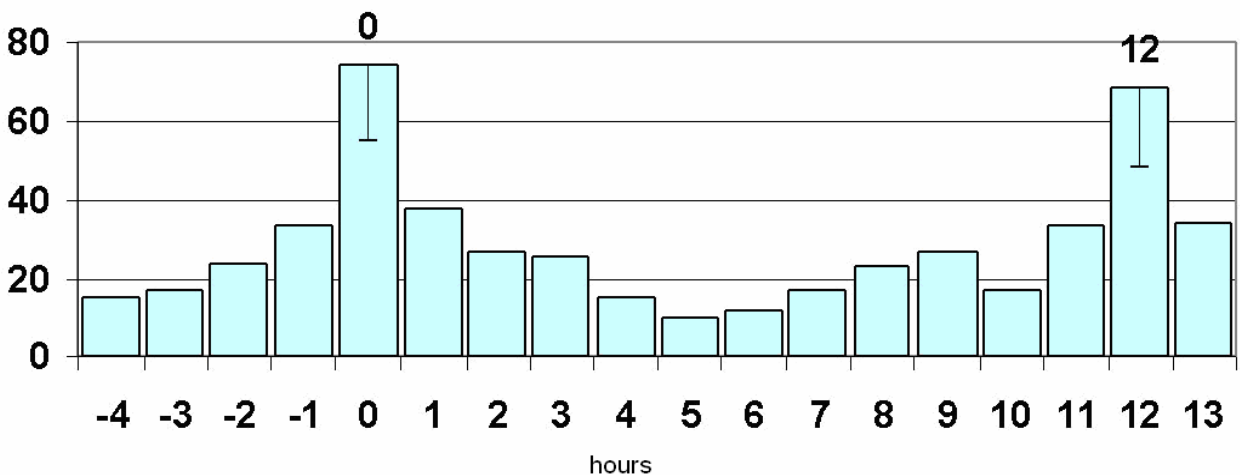

**Fig. 6. Similar histograms are realized with a high likelihood exactly in two years and in 2 years plus 12 hours (with resolution of one hour).** Measurements in August – September, 2000 and

2002. **Abscissa is time interval in hours minus two years. Ordinate is number of similar histogram pairs corresponding to the given size of an interval. Lower 95% confidence limits are marked for peaks.**

At more detailed analysis of two-year interval, when comparing one-minute histograms, it has been shown, that the first "calendar" period is in 2 minutes less then two-year interval and is equal to 1051198, instead of 1051200 minutes; the second "sidereal" period appeared to be in 369 x 2 = 738 minutes longer that two-year interval. It is shown at fig. 7. This result also is reproduced repeatedly.

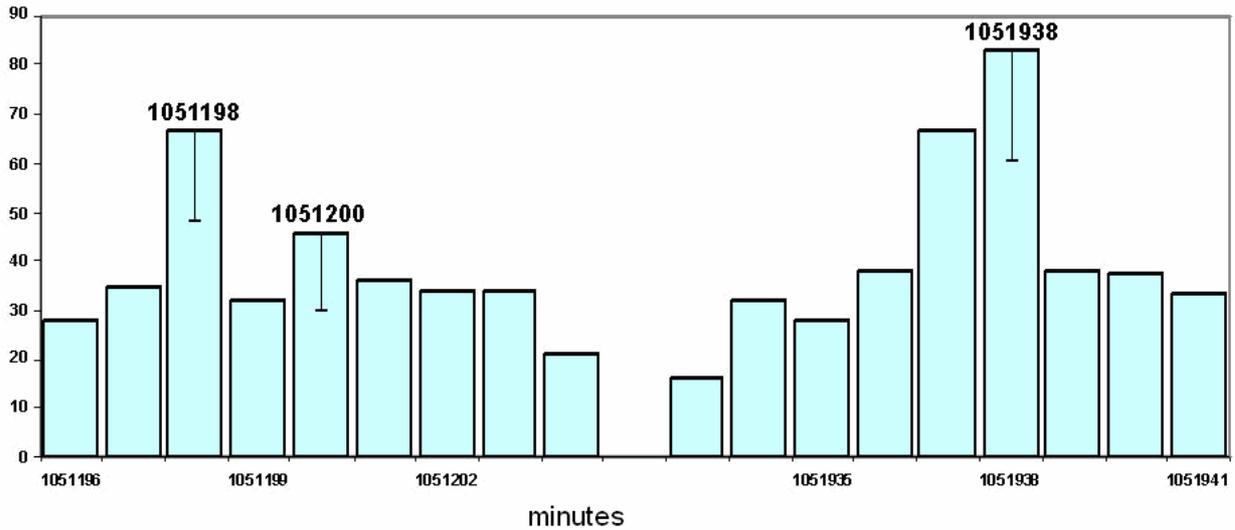

**Fig. 7. For measurements with one-minute resolution, in two-year interval similar histograms are realized with two periods: first is two minutes less then exact two years, and second is 738 minutes more then two years.** Measurements at April, 20, 2001 and 2003. Abscissa is time interval in minutes. Ordinate is number of similar histogram pairs corresponding to the given size of an interval. Lower 95% confidence limits are marked for peaks.

In this case for assurance of higher statistical reliability of defining of size of "calendar" period for two-year interval, comparison of more than 80 000 pairs histograms has been made. (81900 comparisons were made, 3789 or 4,7 % similar pairs were found, and interval distributions of the same shape as at the fig.7 were obtained.)

For tree-year interval with 1-hour histograms there were received two periods again: "calendar", equal to number of hours in three years, and "sidereal", bigger then the first one for 18 hours. It is shown at fig. 8.

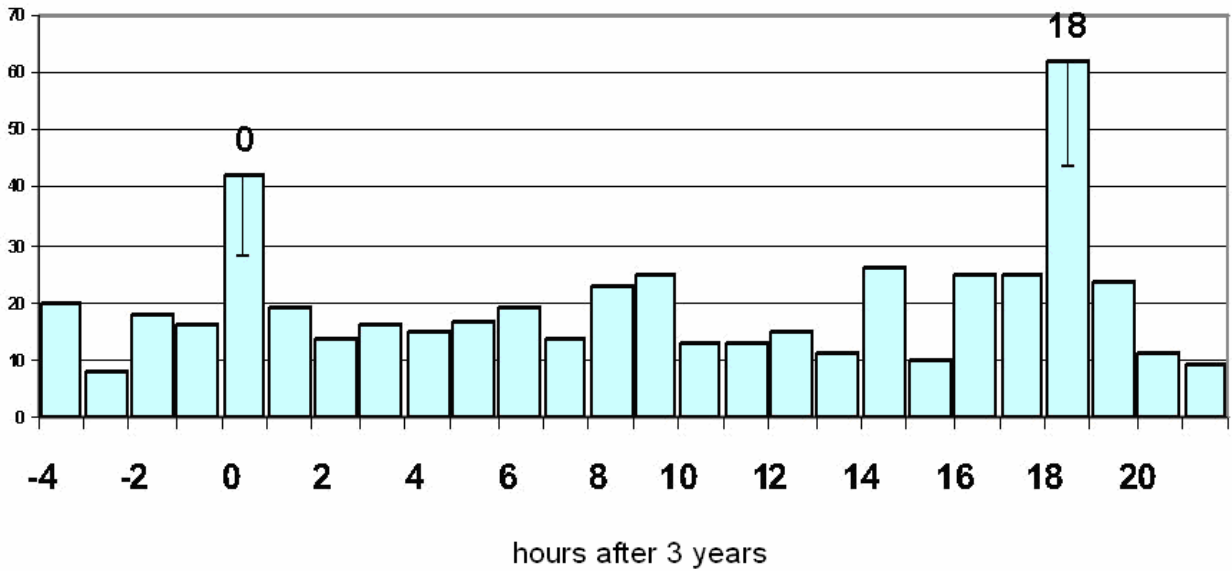

**Fig. 8. Similar histograms are realized exactly in three years and in 3 years plus 18 hours (with resolution of one hour). Measurements in August – October of 2000 and 2003. Abscissa is time interval in hours minus three years. Ordinate is number of similar histogram pairs corresponding to the given size of an interval. Lower 95% confidence limits are marked for peaks.**

To statistically significant determine a value of "calendar" period in three years with the resolution 1 minute, there were performed about 200000 comparisons of histograms. Results of these comparisons are presented at fig. 9.

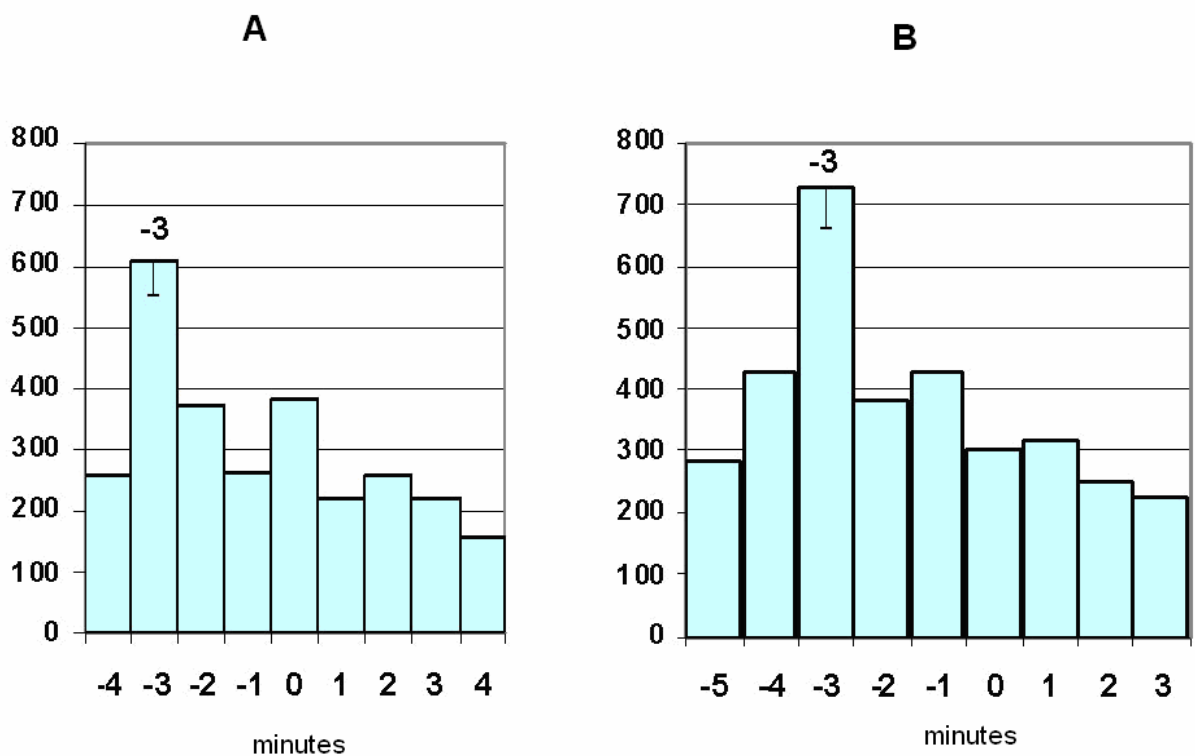

**Fig. 9. In measurements with tree-year interval (in 1576800 minutes) maximum of likelihood of similar histograms is realized at three minutes earlier then exact "calendar" 3-year period. (Measurements in 2000 and 2003: A – in October, Б – in August and November). Abscissa is time interval in minutes minus number of minutes in three years. Ordinate is number of similar histogram pairs corresponding to the given size of an interval. Lower 95% confidence limits are marked for peaks.**

For 4-year interval the tendency discussed above was kept. While not taking into account the leap day (not adding February 29), the "calendar" extremum is observed at 4 minutes earlier then an estimated time. The addition of leap day changes the picture. Now the advancing in time of occurrence of "calendar" extremum makes 8 minutes. Besides, there is a precise "zero" extremum, which is exactly equal to the number of minutes in 4 years (3 regular years (365x3) + leap year (366 days) => [365 x 3 + 366] x 1440 = 2103840 minutes).

In 4 years the shift of "sidereal" extremum without addition of leap day would make 369 x 4 = 1440+36 minutes. After addition of leap day, part of the shift equal to 1440 minutes is compensated. The rest of 36 minutes appears not compensated.

All this was shown at comparison of the histograms constructed by results of one-second measurements on August, 2-3 and on August, 29-30, in 2000 and 2004.

It is shown at fig. 10.

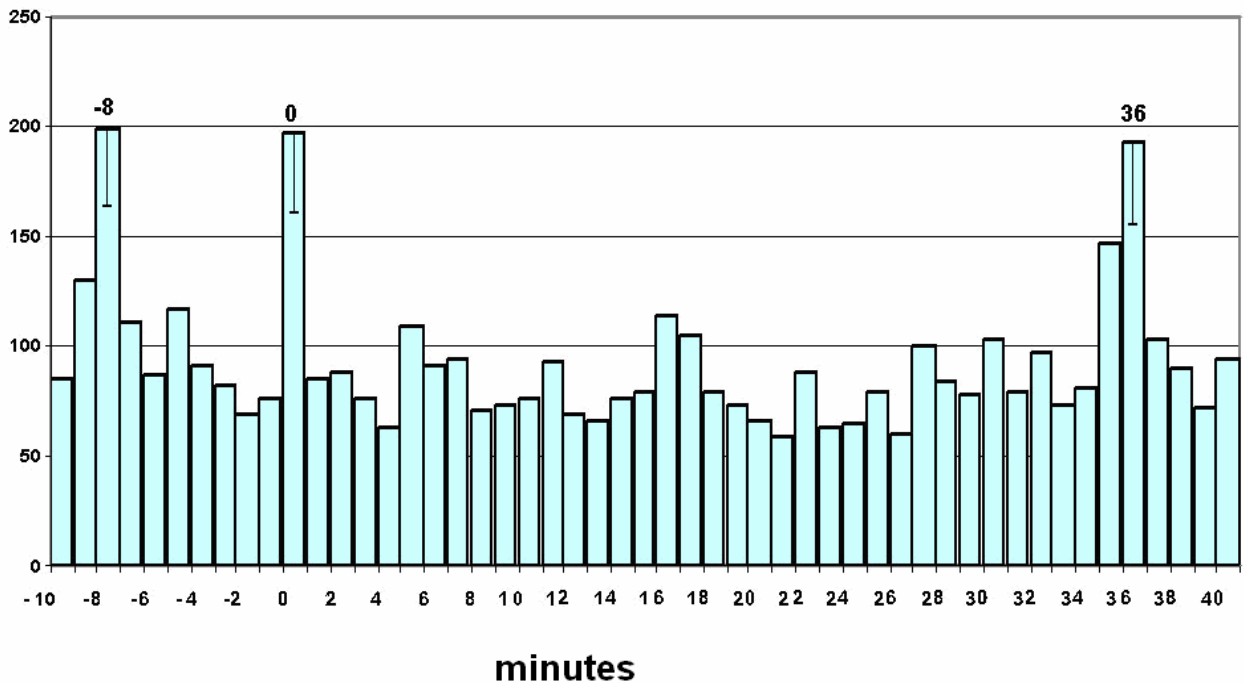

**Fig. 10. In 4 years (measurements in the same dates and time of day in 2000 and 2004, August, 2-3 and 29-30) three extrema of increase in likelihood of occurrence of similar one-minute histograms are observed: first is "calendar" period, 8 minutes before an estimated time, the second is equal to 4 years exactly ("zero point"), and the third "sidereal" is in 36 minutes after "zero point".** Abscissa is time interval in minutes minus number of minutes in three years. Ordinate is number of similar histogram pairs corresponding to the given size of an interval. Lower 95% confidence limits are marked for peaks.

The results presented at fig. 10 can be explained by the following way: *) as known, sidereal day is at 3 minutes and 55.90656 seconds (= 235.90656 seconds) shorter then solar day. In 365 days this difference will make 235.90656 × 365 = 86105.8944 seconds. It is less than duration of average solar day by 58.199 seconds. It means, that for one-year interval the celestial sphere is reproduced approximately 1 minute later then the solar "New Year". We observe annual shift of "calendar" extremum in one minute earlier then solar one. Consequently, the true annual shift of "calendar" extremum is equal to 2 minutes in a year: one minute is compensated by the specified difference of solar and sidereal day. For 4-years interval shift of "calendar" extremum is equal to 8 minutes, as it was shown after taking into account of additional leap date of February 29, compensating an inequality of star and solar day. The display of a "zero extremum", a sharp increase in likelihood of repeated realization of similar histograms, appeared exactly in 4 years, demands additional investigation. The extremum of "36 minutes" according to above-mentioned, is the rest of shift of "sidereal" extremum after compensation of 1440 minutes by inclusion of one leap day, February 29.

*) We are grateful to T.A.Zenchenko and D.P.Harakoz for fairly valuable discussion of these results

---

**Discussion**

During many years, while investigating the fine structure of statistical distributions of results of measurements of processes of the different nature – shapes of corresponding histograms – we discussed the possible nature of the found regularities [1-4]. Not repeating above-mentioned, it is necessary to emphasize major statements.

1. Regular transformation of fine structure of histograms does not contradict to "basics of science".

During last centuries the random variability of results of measurements was, as a rule, only undesirable attribute complicating certain conclusions about the investigated phenomenon. Methods of statistical processing allow minimizing undesirable effects of results variation. On the other hand, integrated characteristics of this variation have allowed drawing the important conclusions on the nature of investigated processes. An example to this can be an evidence of compliance of processes of radioactive decay with Poisson distribution, confirming independence of decay of each atom [3].

The evidence of the regular nature of variations of thin structure of histograms, including "obviously random" processes such as radioactive decay, does not contradict the available data. Modern "criteria of agreement" of hypotheses are integrated and "do not detect" this fine structure [4].

2. Regular changes of fine structure of distributions (the shapes of histograms) are not connected to changes of the measured value itself (its average) and do not depend on process nature. During each given moment, in the given place they with high likelihood are similar at measurements of any processes – from biochemical reactions through radioactive decay [1-4,9].

3. In this connection over 25 years ago we have chosen radioactive decay as the basic object of research. This choice has been proved by apparent independence of radioactive decay of changes of laboratory conditions of carrying out the experiment. We have chosen alpha decay from different kinds of radioactive decay, because of its independence of conditions of measurements and presence of the perfect (semi-conductor) detectors practically excluding instrumental deviations [3,4]. During long-term, whenever possible continuous round-the-clock measurements, we accumulated tons of results of measurements in computer archive.

4. The only common thing at investigation of processes of various natures was a place and time of conducting measurements. It gave us an assumption that the possible reason of the observable phenomena is fluctuations of space-time continuum. Fluctuations of space-time are connected with non-homogeneous mass distribution in surrounding space and moving of measured objects regarding these heterogeneous masses. Such movements are caused by rotation of the Earth, movement of the Earth along the solar orbit, nutations of terrestrial axis, movement of Solar system in the Galaxy, etc. These movements are revealed in changes of interposition of the Earth, the Sun, the Moon and, probably, other space objects.

5. Abovementioned corresponds to results of measurements in different geographical points from Arctic through Antarctic Region, at ground laboratories in the different countries and on board of the ships at Pacific, Indian, Atlantic and Arctic oceans [5,10].

The major result of these measurements is detection of synchronism of changes of the shape of histograms in different geographical points at the same local time.

6. The establishment of equality of the daily period of change of likelihood of repeated occurrence of histograms of the given shape, to 1440 minutes ("solar day"), as well as to 1436 minutes ("sidereal day") [4] was essentially important.

7. Presence of the daily period in changes of the shape of histograms, means, that "the affecting cause" is shielded by the Earth and appears above the horizon in the process of the Earth rotation. The narrowness of corresponding extrema, their one-minute resolution, confirms sharp anisotropy of this "cause". In direct experiments with collimators, restricting the flow of alpha particles in certain directions, this sharp anisotropy has received accurate confirmation [6,7.] Particularly convincing picture has been obtained in experiments with rotated collimators: the shape of histograms depends on a direction of alpha particle flow [11].

8. The data presented in this paper on the nature of the yearly periods in variations of the histogram shapes, open new features of the investigated phenomenon. It confirms the dependence of the histogram shape on exposition relative to the celestial sphere and to the Sun.

As appeared, in a year, histograms of the given shape show again in 2 minutes ahead of time of corresponding orientation of the Earth and the Sun. Such shift of "calendar" extremum can be explained either by nutation of the Earth axis, or by movement of Solar system, as the whole, regarding other objects in the Galaxy.

The "sidereal" extremum corresponds to the position of the Earth in a circumsolar orbit, when the Sun appears (is projected on) against the same star, as one year ago. It remains not clear why this extremum is not shifted year after year as well as "calendar" extremum.

9. The fine structure of analyzed histograms, presence of the "permitted" and "forbidden" values of measurements, narrowness of corresponding "peaks" and "hollows", rather reminds the phenomena of an interference. This structure cannot be explained by presence of several likelihood constants ("constants of disintegration") [4]. Interference is property of wave processes. Therefore it suggests a possible wave nature of the investigated phenomenon. The similar explanation of the nature of the paradox of experiments with collimators suggested by D.P.Harakoz seems plausible. Collimators that we use, "cut out" sites about 5 angular degrees on celestial sphere, which correspond to 20 minutes of time at rotation of the Earth about the axis. We see (at figures presented in this paper) extrema with the resolution of one minute. Such sharp peaks are most naturally explained by interference – summation of wave streams coming in different angles, or distinguished by phase. The nature of these wave streams is a subject for discussion.

--ooOoo--

In conclusion it is necessary to note, that discussed phenomena are rather complex and are defined, apparently, by set of factors. In this connection various hypotheses were stated by authors of papers [12-16].

## Bibliography

[1] Shnoll S.E. , Kolombet V.A. , Pozharski E.V. , Zenchenko T.A., Zvereva I.M. and Konradov A.A. :**1998**, Realization of discrete states during fluctuations in macroscopic processes, *Physics-Uspehi* **162**(10), 1129-1140.

[2] Shnoll S.E. , Pozharski E.V. , Zenchenko T.A. , Kolombet V.A. , Zvereva I.M. and Konradov A.A.: **1999**, Fine structure of distributions in measurements of different processes as affected by geophysical and cosmophysical factors,  *Phys.& Chem. Earth A: Solid Earth & Geod.***24**(8), 711–714.

[3] Shnoll S.E. , Zenchenko T.A. , Zenchenko K.I., Pozharski E.V., Kolombet V.A., and Konradov A.A.: **2000**, Regular variation of the fine structure of statistical distributions as a consequence of cosmophysical agents, *Physics-Uspehi* **43**(2), 205-209

[4] Shnoll,S.E.: **2001**, Discrete distribution patterns: arithmetic and cosmophysical origins of their macroscopic fluctuations, *Biophysics* **46**(5),733-741.

[5]Shnoll S.E., Rubinstein I.A., Zenchenko K.I. , Zenchenko T.A., Konradov A.A , Shapovalov S.N., Makarevich A.V., Gorshkov E.S., and Troshichev O.A.: **2003**, Dependence of

"Macroscopic Fluctuations" on Geographic Coordinates (by Materials of Arctic and Antarctic Expeditions) *Biophysics* **48**(5), -1123-1131

[6] Simon E. Shnoll, Konstantin I. Zenchenko, Iosas I. Berulis, Natalia V. Udaltsova, Serge S. Zhirkov and Ilia A. Rubinstein, **2004**, , Dependence of "Macroscopic Fluctuations" on Cosmophysical Factors. Spatial Anisotropy. *Biophysics* **49**(1), -129 -139

[7] Simon E. Shnoll, Konstantin I. Zenchenko, Iosas I. Berulis, Natalia V. Udaltsova and Ilia A. Rubinstein , **2004** Fine structure of histograms of alpha-activity measurements depends on direction of alpha particles flow and the Earth rotation: experiments with collimators http://arxiv.org/abs/physics/0412007

[ 8] S.E. Shnoll, V.A. Kolombet, , Zenchenko T.A., Pozharskii E.V., Zvereva I.M. and Konradov A.A., **1998**, Cosmophysical Origin of "Macroscopic Fluctations" *Biophysics*(in russ.), **43**(5),909-915

[9] Fedorov M.V., Belousov L.V., Voeikov V.L., Zenchenko T.A., Zenchenko K.I., Pozharskii E.V., Konradov A.A.and Shnoll S.E.: **2003**, Synchronous changes in dark current fluctuations in two separate photomultipliers in relation to Earth rotation, *Astrophysics and Space Science.* **283:**3-10.

[10] S. E. Shnoll, K. I. Zenchenko, S. N. Shapovalov,E. S. Gorshkov, A. V. Makarevich and O. A. Troshichev.:2004 The specific form of histograms presenting the distribution of data of α-decay measurements appears simultaneously in the moment of New Moon in different points from Arctic to Antarctic.**2004** http://arxiv.org/abs/physics/0412152

[11]. S. E. Shnoll, I.A.Rubinshtejn, K. I. Zenchenko, V.A.Shlekhtarev, A.V.Kaminsky, A.A.Konradov, N.V.Udaltsova.:**2005**, Experiments with rotating collimators cutting out pencil of alpha-particles at radioactive decay of $^{239}$Pu evidence sharp anisotropy of space . http://arxiv.org/abs/physics/0501004

[12]. A.A.Kirillov and K.I.Zenchenko: **2001** On the Probability of Disturbance of the Poisson Statistics in Processes of Radioactive Decay Type, Biophysics Vol **46**,(5), 841-849

[13].V.K.Lyapidevskii, **2001**: Diurnal Variations in the Flux of Alpha-particles as Possible Evidence for Changes in the Vector of Velocity of Movement of an Experimental Set-up relative to a Relic System. Biophysics Vol **46**,(5), 850-851

[14].I.M.Dmitrievskii., **2001**: A Possible Explanation of the Phenomenon of Cosmophysical Fluctuations. Biophysics Vol **46**,(5), 885-855

[15].V.A.Namiot, **2001**: On the Theory of the Effect of "Macroscopic Fluctuations", Biophysics Vol **46**,(5), 856-858

[16].L.A.Blumenfeld and T.A.Zenchenko, **2001** : Quantum Transitions between States and Cosmophysical Fluctuations, Biophysics Vol **46**,(5), 859-861

**ACKNOWLEDGEMENTS**

This work is dedicated to the memory of A.A. Konradov. His friendly support and participation in these investigations created a cheerful atmosphere that helped us to view and overcome unavoidable difficulties optimistically.

S.E.Shnoll is indebted to M.N. Kondrashova and L.A. Blumenfeld for mutual understanding.

The authors are grateful to our colleagues V.A.Kolombet, T.A. Zenchenko, I.A. Rubinshtein, and D.P. Kharakoz for many years of collaboration and valuable discussions.

The authors are grateful to V.A.Namiot, O.A.Troshichev, E.S.Gorshkov, S.N.Shapovalov, B.M.Vladimirsky, B.V. Komberg, V. K. Lyapidevskii, I.M.Dmitrievskii, A.A.Kirilov, D.S.Chernavskii, A.V.Kaminskii for valuable discussions.

The vivid interest of V.P.Tikhonov to the problem studied and his generous financial support were essential.

Correspondence and requests for materials should be addressed to S.E.S.

(e-mail: shnoll@iteb.ru).